\documentstyle[preprint,aps]{revtex}
\tighten
\draft
\widetext
\input epsf
\preprint{HUTP-98/A044, NUB 3180}
\bigskip
\bigskip
\begin{document}
\title{A Three-Family $SU(4)_c \otimes SU(2)_w \otimes U(1)$ Type I Vacuum}
\medskip
\author{Zurab Kakushadze\footnote{E-mail: 
zurab@string.harvard.edu}}
\bigskip
\address{Lyman Laboratory of Physics, Harvard University, Cambridge, 
MA 02138\\
and\\
Department of Physics, Northeastern University, Boston, MA 02115}
\date{June 5, 1998}
\bigskip
\medskip
\maketitle

\begin{abstract}
{}We construct a four dimensional chiral ${\cal N}=1$ space-time supersymmetric
perturbative Type I vacuum corresponding to a compactification on a toroidal ${\bf Z}_2\otimes {\bf Z}_2 \otimes {\bf Z}_3$ orbifold with a discrete Wilson line. 
This model is non-perturbative from the heterotic viewpoint. It has three chiral families in the $SU(4)_c\otimes SU(2)_w\otimes U(1)$ subgroup of the total gauge group. 
We compute the tree-level superpotential in this model. 
There appears to be no obvious obstruction to Higgs the gauge group down to $SU(3)_c\otimes SU(2)_w\otimes U(1)_Y$ and obtain the Standard Model gauge group with three chiral families.
\end{abstract}

{}One of the outstanding questions in string theory is how does it describe {\em our} universe.
This is a difficult question to answer for the classical string vacua have a very large degeneracy.
Moreover, this degeneracy persists to all loops in perturbation theory. It is, however, reasonable to expect that non-perturbative effects might (at least partially) lift the degeneracy of superstring vacua. Non-perturbative dynamics should also be responsible for supersymmetry breaking as
supersymmetry is believed to be unbroken perturbatively in superstring vacua. Understanding of non-perturbative effects in superstring theory is therefore of utmost importance.

{}In recent years much progress has been made in understanding non-perturbative string dynamics. In particular, various consistent superstring theories (which are distinct perturbatively) are now believed to be different manifestations (in the appropriate regimes)
of an underlying unified theory. Thus, the (conjectured) web of dualities between different string theories (often) allows to map non-perturbative phenomena in one theory to perturbative phenomena in another theory. In particular, some superstring vacua inaccessible perturbatively in one framework may have perturbative realizations in a dual picture.
  
{}Phenomenologically oriented model building in superstring theory for many years has been mostly confined to the perturbative heterotic superstring framework where it was facilitated by the existence of relatively simple rules. In particular, perturbative heterotic superstring enjoys the constraints of conformal field theory and modular invariance which serve as guiding principles for model building. Given the success of string dualities in shedding light on non-perturbative string dynamics, it is natural to attempt construction of phenomenologically interesting superstring vacua which would be non-perturbative from the heterotic viewpoint. This, in particular, could {\em a priori} provide us with clues for solving some of the phenomenological problems encountered in heterotic model building.  

{}From the above viewpoint, one of the promising directions appears to be studying four dimensional Type I compactifications in the phenomenological context. One of the reasons for such a belief stems from the conjectured Type I-heterotic duality \cite{PW}. In particular, certain non-perturbative effects on the heterotic side (such as dynamics of NS 5-branes which is difficult to study from the heterotic viewpoint) are mapped to perturbative Type I effects ({\em e.g.}, D5-branes are the Type I duals of heterotic NS 5-branes) which are under much better control.

{}The simplest compactifications of Type I are those on toroidal orbifolds. Such Type I vacua
can be viewed as Type IIB orientifolds with a certain choice of the orientifold projection.
Orientifolds are generalized orbifolds 
that involve world-sheet parity reversal along with geometric symmetries 
of the theory \cite{group}. Orientifolding procedure results in an unoriented closed string theory.
Consistency then generically requires introducing open strings that can 
be viewed as starting and ending on D-branes \cite{Db}.  
Global Chan-Paton charges
associated with D-branes manifest themselves as a gauge symmetry in 
space-time. The orientifold techniques have been successfully
applied to the construction of six dimensional ${\cal N}=1$ space-time supersymmetric 
orientifolds of Type IIB compactified on orbifold limits of K3 (that is, toroidal orbifolds
$T^4/{\bf Z}_N$, $N=2,3,4,6$) \cite{PS}. 
These orientifold models generically contain more than one tensor multiplet and/or enhanced gauge symmetries from D5-branes in their massless spectra,
and, therefore, describe six dimensional vacua which are non-perturbative from the 
heterotic viewpoint.   

{}The orientifold construction has subsequently been generalized to four dimensional ${\cal N}=1$ space-time supersymmetric compactifications. Several
such orientifolds have been constructed \cite{BL,Sagnotti,KS1,su6,class}. Some of these models, namely, the ${\bf Z}_3$ \cite{Sagnotti}, ${\bf Z}_7$, ${\bf Z}_3\otimes {\bf Z}_3$ \cite{KS1}, and 
$\Delta(3\cdot 3^2)$ (the latter group is non-Abelian) \cite{class} orbifold
models have perturbative heterotic duals \cite{Sagnotti,KS1,class}. Others, such as the ${\bf Z}_2\otimes {\bf Z}_2$ \cite{BL}, ${\bf Z}_6$ \cite{KS1} and ${\bf Z}_2\otimes {\bf Z}_2\otimes {\bf Z}_3$ \cite{su6} orbifold models are non-perturbative
from the heterotic viewpoint as they contain D5-branes.  
In particular, some of these models (such as the ${\bf Z}_6$ \cite{KS1} and ${\bf Z}_2\otimes {\bf Z}_2\otimes {\bf Z}_3$ \cite{su6} orbifold models) are examples of consistent {\em chiral} 
${\cal N}=1$ string vacua in four dimensions that are non-perturbative from the heterotic viewpoint.

{}The above seven orbifold groups have been recently argued in \cite{KST,zura1,class} to lead to {\em perturbatively} well defined orientifolds. In particular,
in \cite{KST} conditions necessary for world-sheet
consistency of six and four dimensional ${\cal N}=1$ supersymmetric Type IIB 
orientifolds were studied. Moreover, it was also argued 
in \cite{KST,zura1,class}
that for all the other {\em a priori} consistent choices (including those discussed in 
\cite{Zw,Iba}) of the orbifold group $\Gamma$ (such that
$T^6/\Gamma$ has $SU(3)$ holonomy) the corresponding
orientifolds contain sectors
which are non-perturbative ({\em i.e.}, these sectors have no world-sheet description). These
sectors can be thought of as arising from D-branes wrapping various (collapsed) 2-cycles
in the orbifold. 

{}Based on the results in \cite{KST,zura1}, in \cite{class} a classification of perturbative (from the orientifold viewpoint) four dimensional ${\cal N}=1$ Type I compactifications (on toroidal orbifolds $T^6/\Gamma$) has been given. In particular, all such models, including those with non-trivial NS-NS
antisymmetric tensor backgrounds, were explicitly constructed in \cite{class}. All of these models, however, have been constructed with the assumption of no non-trivial Wilson lines.
Inclusion of non-trivial Wilson lines (severely constrained by the requirement that they be compatible with the orbifold group) in these models is completely straightforward. In particular, turning on a {\em discrete} Wilson line can be thought of as a freely acting orbifold which amounts to 
appropriately shifting the momentum plus winding lattice corresponding to the compactified six
dimensions, that is, it is a shift in the six-torus $T^6$. The action of this orbifold group on the Chan-Paton factors (constrained by the requirement of tadpole cancellation) can be non-trivial and result in further breaking of gauge symmetry. The massless spectrum then is obtained by considering the corresponding orbifold projection of the parent theory (without the Wilson line)
onto the states invariant under the action of the orbifold group.

{}The classification of \cite{class} is useful in a sense that, although turning on Wilson lines
is rather straightforward (as opposed to, for instance, turning on non-zero NS-NS antisymmetric tensor backgrounds studied in \cite{class}), it allows us to zoom onto the orbifolds most promising from the phenomenological viewpoint. By viewing the spectra of the Type I vacua explicitly constructed in \cite{class}, it becomes clear that one of the models that stand out 
from the phenomenological perspective is the ${\bf Z}_2\otimes{\bf Z}_2 \otimes {\bf Z}_3$ orbifold model (with no NS-NS antisymmetric tensor background) originally constructed in \cite{su6}. This model has an $SU(6)\otimes Sp(4)$ gauge subgroup, and there are
three chiral families of $SU(6)$. However, as was already pointed out in \cite{su6}, the matter content in this model is such that it is impossible to break the $SU(6)$ gauge subgroup down to the Standard Model gauge group $SU(3)_c \otimes SU(2)_w \otimes U(1)_Y$ by Higgsing. 
Nonetheless, there is another possibility. Namely, in the following we will see that the $SU(6)\otimes Sp(4)$ gauge group in this model can be broken by a {\em discrete} Wilson line to yield a model with three families and a phenomenologically acceptable gauge group $SU(4)_c\otimes SU(2)_w \otimes U(1)$ (which can further be broken down to $SU(3)_c\otimes SU(2)_w \otimes U(1)_Y$ by Higgsing).   

{}Next, we turn to the explicit construction of the three-family $SU(4)_c\otimes SU(2)_w \otimes U(1)$ Type I model mentioned above. Consider a Type I compactification on the toroidal orbifold ${\cal M}=T^6/\Gamma$ with zero NS-NS $B$-field (that is, the internal components of the $B$-field are all zero: $B_{ij}=0$, $i,j=1,\dots,6$). Here $\Gamma\approx{\bf Z}_2\otimes {\bf Z} _2\otimes {\bf Z}_3$ is the orbifold group. Let $g$, $R_1$ and $R_2$ be the generators of the ${\bf Z}_3$ and the two ${\bf Z}_2$ subgroups of $\Gamma$. Then the action of $g$ and $R_s$
($R_3=R_1R_2$) on the complex coordinates $z_{s^\prime}$ are given by (here for the sake of simplicity we can assume that the six-torus $T^6$ factorizes into a product $T^6=T^2\otimes T^2\otimes T^2$ of three two-tori, and the complex coordinates $z_s$ ($s=1,2,3$) parametrize these three two-tori): 
\begin{equation}
 g z_s=\omega z_s~,~~~R_s z_{s^\prime}=-(-1)^{\delta_{ss^\prime}} z_{s^\prime}~.
\end{equation} 
(There is no summation over the repeated indices here.) Here $\omega=\exp(2\pi i/3)$. The Calabi-Yau three-fold ${\cal M}=T^6/\Gamma$ (whose holonomy is $SU(3)$) has 
Hodge numbers $(h^{1,1},h^{2,1})=(36,0)$. Thus, the closed string sector of Type I on ${\cal M}$ consists of ${\cal N}=1$ supergravity, the dilaton plus axion supermultiplet, and 36 chiral supermultiplets neutral under the open string sector gauge group\footnote{Some of these supermultiplets transform non-trivially under the anomalous $U(1)$'s in the open string sector, however.}.

{}The Type I model we are discussing here can be viewed as the $\Omega$ orientifold of Type II
B compactified on ${\cal M}$, where $\Omega$ is the world-sheet parity reversal. The tadpole cancellation conditions require introducing 32 D9-branes (corresponding to the $\Omega$ element of the orientifold group) as well as three sets of D5-branes (which we will refer to as D$5_s$-branes) with 32 D5-branes in each set. Moreover, the action of the orbifold group $\Gamma$ on the Chan-Paton charges carried by the D9- and D$5_s$-branes is described by
the corresponding Chan-Paton matrices $\gamma_{g_a}$ where $g_a$ are the elements of the
orbifold group $\Gamma$. These matrices must form a projective representation of (the double cover of) the orbifold group $\Gamma$. The twisted tadpole cancellation conditions give constraints on traces of the Chan-Paton matrices. In particular, we have the following tadpole cancellation conditions (here we note that the orientifold projection $\Omega$ on the D9-brane Chan-Paton charges is required to be of the $SO$ type in this model) \cite{su6}:
\begin{equation}
 {\mbox{Tr}}(\gamma_g)=-2~,~~~
 {\mbox{Tr}}(\gamma_{R_s})={\mbox{Tr}}(\gamma_{gR_s})=0~.
\end{equation}   
Here we are using the convention (which amounts to not counting orientifold images of D-branes) where the Chan-Paton matrices are $16\times 16$ matrices (and not $32\times 32$ matrices which would be the case if we counted both the D-branes and their orientifold images).
The solution (up to equivalent representations) to the twisted tadpole cancellation conditions is given by \cite{su6}:
\begin{equation}
 \gamma_{g,9}={\mbox{diag}}({\bf W}\otimes {\bf I}_3, {\bf I}_4)~,~~~
 \gamma_{R_s,9}=i\sigma_s \otimes {\bf I}_8~.
\end{equation}
Here ${\bf W}={\mbox{diag}}(\omega,\omega,\omega^{-1},\omega^{-1})$, $\sigma_s$ are the Pauli matrices, and ${\bf I}_n$ is an $n\times n$ identity matrix.
(The action on the D$5_s$-branes is similar.) The massless spectrum of this model was worked out in \cite{su6}. The gauge group is $[U(6)\otimes Sp(4)]_{99}\otimes \bigotimes_{s=1}^3 [U(6)\otimes Sp(4)]_{5_s 5_s}$. (Here we are using the convention where $Sp(2N)$ has rank $N$.) There are three chiral generations in each $SU(6)$ subgroup in this model.

{}Next, we would like to turn on a {\em discrete} Wilson line in this model such that it would 
break the gauge group in a phenomenologically favorable fashion. Such a Wilson line is not difficult to find. Thus, consider a freely acting orbifold which amounts to a ${\bf Z}_3$ shift in the
third $T^2$ (parametrized by the complex coordinate $z_3$). That is, let this two-torus be defined by the identifications $z_3\sim z_3+n^\alpha e_\alpha$, where $n^\alpha\in {\bf Z}$, and $e_\alpha$ ($\alpha=1,2$) are the constant vielbeins. Then this ${\bf Z}_3$ shift $S$ has the following action on $T^2$: $Sz_3=z_3+{1\over 3} m^\alpha e_\alpha$
for some integers $m^\alpha$ such that ${1\over 3} m^\alpha e_\alpha\not\in \Lambda$, where
$\Lambda=\{n^\alpha e_\alpha\}$ is the lattice defining this torus (that is, $T^2={\bf C}/\Lambda$). This shift commutes with the ${\bf Z}_3$ twist generated by the orbifold group
element $g$, and also with the ${\bf Z}_2$ twist generated by the orbifold group
element $R_3$. However, it does not commute with the elements $R_1$ and $R_2$. In particular, $S$ and $R_1$ generate a non-Abelian group isomorphic to $D_3$. Similarly, 
$S$ and $R_2$ also generate a non-Abelian group isomorphic to $D_3$. The action of the ${\bf Z}_3$ orbifold group generated by $S$ on the Chan-Paton factors can be chosen as follows. First, it can be trivial: $\gamma_{S,9}=\gamma_{S,5_s}={\bf I}_{16}$. In this case the effect of this freely acting orbifold is simply to rescale the radii of the third $T^2$. In particular, the open string sector is unchanged. (This, in particular, is consistent with the tadpole cancellation conditions.) Here we are interested in a non-trivial action of $S$ on the Chan-Paton factors. 
This action is subject to the corresponding tadpole cancellation conditions.
There is no restriction on the trace of $\gamma_S$ itself. (This can be seen, for instance,  by noting that in the $S$ and $S^{-1}$ ``twisted'' closed string sectors there are no massless states.)
However, there are constraints on the traces of $\gamma_{S^k g^{k^\prime}}$ ($k,k^\prime=1,2$):
\begin{equation}
 {\mbox{Tr}}(\gamma_{S^k  
 g^{k^\prime},9})={\mbox{Tr}}(\gamma_{g,9})={\mbox{Tr}}(\gamma_{g^{-1},9})=-2~.
\end{equation}  
(Similar constraints apply to the D$5_s$-brane Chan-Paton matrices as well.) An appropriate solution to these constraints reads (and similar expressions hold for the D$5_s$-branes):
\begin{equation}
 \gamma_{S,9}={\mbox{diag}}({\bf W}^\prime\otimes {\bf I}_{2n}, 
 {\bf I}_{12-4n}, {\bf W}^\prime \otimes {\bf I}_n, {\bf I}_{4-2n})~,
\end{equation}
where ${\bf W}^\prime ={\mbox{diag}}(\omega,\omega^{-1})$. Here $n=1,2$. Note that $\gamma_{S,9}$ commutes with $\gamma_{R_3,9}$, and generates
$D_3$ groups with $\gamma_{R_1,9}$ and $\gamma_{R_2,9}$ as it should from our previous discussions.

{}For $n=2$ the resulting gauge group is $[U(2)^4 ]_{99}\otimes \bigotimes_{s=1}^3 
[U(2)^4 ]_{5_s 5_s}$ which does not contain an $SU(3)_c$ subgroup so we will not be interested in this possibility here. However, for $n=1$ we have the following gauge group:
$[U(4)\otimes SU(2)\otimes U(1)^3]_{99}\otimes \bigotimes_{s=1}^3 [U(4)\otimes SU(2)\otimes U(1)^3]_{5_s 5_s}$. (Here we have used the fact that $Sp(2)$ is isomorphic to $SU(2)$.) This is a phenomenologically acceptable gauge group as it contains the Standard Model gauge group $SU(3)_c\otimes SU(2)_w \otimes U(1)_Y$ as a subgroup. The massless open string spectrum 
of this model is given in Table I.

{}Note that the massless spectrum of this model is free of non-Abelian gauge anomalies. 
This model might be phenomenologically interesting for the following reason. Suppose the gauge group of the observable world comes solely from one type of D5-branes, say, D$5_3$-branes.
We can identify the $SU(4)$ subgroup of $[U(4)\otimes SU(2)\otimes U(1)^3]_{5_3 5_3}$ with
$SU(4)_c$ corresponding to strong interactions (assuming that this $SU(4)_c$ contains the
color subgroup $SU(3)_c$ of QCD), and the $SU(2)$ subgroup of $[U(4)\otimes SU(2)\otimes U(1)^3]_{5_3 5_3}$ with $SU(2)_w$ corresponding to the weak interactions. At some point we would need to break $SU(4)_c$ down to $SU(3)_c$. This could happen as follows. It is not difficult to see (taking into account the superpotential which we give in a moment) that, say, 99 gauge group can be Higgsed completely. Then one could break $SU(4)_c$ in the $5_3 5_3$ sector to $SU(3)_c$ by giving appropriate vevs to the fields $P^3$ and $R^3$. (We should mention, however, that determining whether such a scenario can indeed be realized once the complete dynamics (including possible non-perturbative effects) is taken into account would require much more detailed analyses.) Once $SU(4)_c$ is broken down $SU(3)_c$, one would also need to break the remaining $U(1)$'s so that at the end we are left
with $SU(3)_c\otimes SU(2)_w\otimes U(1)_Y$. It not difficult to see that we then end up with three chiral generations charged under the Standard Model gauge group  
$SU(3)_c\otimes SU(2)_w\otimes U(1)_Y$. (However, one would need to check whether the extra ``vector-like'' matter in the $5_3 5_3$ and other sectors decouples which depends on more detailed dynamics.) In particular, the fields $X^3_k$ ($k=1,2,3$) can be seen to give rise to the three families of left-handed up-quarks (coming from the $({\bf 3},{\bf 2})$ irrep of $SU(3)_c\otimes SU(2)_w$). 

{}A detailed phenomenological analysis of this model 
seems worthwhile but 
is beyond the scope of this paper and will be discussed elsewhere. Here we should mention that many issues 
(such as proton stability, gauge and gravitational coupling unification and all that)
would need to be addressed to see whether this model has a chance to be realistic. Such phenomenological analyses would require the knowledge of the tree level superpotential. Here we give the non-vanishing renormalizable terms in this superpotential (we suppress the actual values of the Yukawa couplings, however):
\begin{eqnarray}
 {\cal W}=&&\epsilon_{ijk} \Phi^s_i X^s_j X^s_k+\epsilon_{ijk} \phi^s_i\chi^s_j
 {\widetilde \chi}^s_k +
 y_{s s^\prime k} \Phi^s_k Q^{ss^\prime} Q^{ss^\prime} + 
 y_{s s^\prime k} \phi^s_k q^{ss^\prime} {\widetilde q}^{ss^\prime} +
 y_{ss^\prime k} X^s_k P^{ss^\prime} R^{ss^\prime}+\nonumber\\
 &&
 y_{ss^\prime k} \chi^s_k p^{ss^\prime} {\widetilde r}^{ss^\prime}+
 y_{ss^\prime k} {\widetilde \chi}^s_k {\widetilde p}^{ss^\prime}  
 {r}^{ss^\prime}+P^{ss^\prime} {Q}^{s^\prime s^{\prime\prime}} R^{s^{\prime\prime} s}+
 p^{ss^\prime} {\widetilde q}^{s^\prime s^{\prime\prime}} r^{s^{\prime\prime} s}+
 {\widetilde p}^{ss^\prime} {q}^{s^\prime s^{\prime\prime}} 
 {\widetilde  r}^{s^{\prime\prime} s}~.
\end{eqnarray}
Here summation over the repeated indices is understood. Also, $y_{ss^\prime\alpha}=\epsilon_{ss^\prime\alpha}$ if $s\not=0$, and 
$y_{0s^\prime\alpha}=\delta_{s^\prime\alpha}$. The condition $s\not=s^\prime\not=s^{\prime\prime}\not=s$ is implicitly assumed in the above expressions, and we are using the compact notation where $P^{s^\prime s}=P^{s s^\prime}$, 
$p^{s^\prime s}={\widetilde p}^{s s^\prime}$, $Q^{s^\prime s}=R^{s s^\prime}$, 
$q^{s^\prime s}=r^{s s^\prime}$,  and
${\widetilde q}^{s^\prime s}={\widetilde r}^{s s^\prime}$, $s,s^\prime,s^{\prime\prime}=0,1,2,3$,
and $s,s^\prime,s^{\prime\prime}=0$ label the D9-branes in the obvious way.

{}Here some remarks are in order. First, turning on additional Wilson lines 
(acting non-trivially on all the Chan-Paton factors)
in this model would break the gauge symmetry too much (and, in fact, the final gauge group would not have an 
$SU(3)_c$ subgroup). Also, turning on Wilson lines in other four dimensional ${\cal N}=1$ models (with or without the $B$-field) classified in \cite{class} does not appear to lead to 
phenomenologically interesting models. There are however some other 
models without Wilson lines with phenomenologically acceptable gauge groups.  
Thus, the ${\bf Z}_6$ orbifold model with a certain non-trivial NS-NS antisymmetric tensor background has an $SU(4)\otimes SU(2)\otimes SU(2)$
gauge subgroup \cite{ST,class}. Basically, there are two possibilities here \cite{ST}. 
In the first case
the $SU(4)\otimes SU(2)\otimes SU(2)$ gauge group comes solely from either the 
99 or 55 sectors.
Then we have only one chiral family plus one vector-like family. In the second case the $SU(2)\otimes SU(2)$ subgroup is a linear combination of the corresponding 99 and 55 gauge groups whereas the $SU(4)$ subgroup comes solely from either the 99 or 55 sectors. In this case
we have 3 chiral families \cite{ST}. However, a more 
careful examination of the superpotential in this
model derived in \cite{class} shows that in this scenario there are no other matter fields charged under the $SU(4)$ subgroup of the Pati-Salam
gauge group $SU(4)\otimes SU(2)\otimes SU(2)$. This makes it impossible to break the latter down to $SU(3)_c\otimes SU(2)_w\otimes U(1)_Y$ (at least in the unbroken supersymmetry limit).

{}This work was supported in part by the grant NSF PHY-96-02074, 
and the DOE 1994 OJI award. I would like to thank Pran Nath and Henry Tye for discussions. 
I would also like to thank Albert and Ribena Yu for 
financial support.

\begin{table}[t]
\begin{tabular}{|c|c|l|}
 Model and Gauge Group& Field & Charged Matter  
  \\
 \hline
   ${\bf Z}_2\otimes{\bf Z}_2\otimes {\bf Z}_3$&
 $\Phi_k$ &
 $3\times [({\bf 6},{\bf 1})(-2,0,0,0)_L]_{99}$
 \\
& $\phi_k$ &
 $3\times [({\bf 1},{{\bf 1}})(0,-1,-1,0)_L]_{99}$  \\
$[U(4)\otimes SU(2)\otimes U(1)^3]_{99}\otimes$&$X_k$ &
 $3\times [({\bf 4},{{\bf 2}})(+1,0,0,0)_L]_{99}$  \\
  $\bigotimes_{s=1}^3 [U(4)\otimes SU(2)\otimes U(1)^3]_{5_s 5_s}$ &$\chi_k$ &
 $3\times [({\bf 1},{{\bf 1}})(0,+1,0,-1)_L]_{99}$  \\ 
&${\widetilde \chi}_k$ &
 $3\times [({\bf 1},{{\bf 1}})(0,0,+1,+1)_L]_{99}$  \\ 
 &$\Phi^s_k$ &
 $3\times [({\bf 6}_s,{\bf 1}_s)(-2_s,0_s,0_s,0_s)_L]_{5_s 5_s}$
 \\
 &$\phi^s_k$ &
 $3\times [({\bf 1}_s,{{\bf 1}}_s)(0_s,-1_s,-1_s,0_s)_L]_{5_s 5_s}$  \\
 &$X^s_k$ &
 $3\times [({\bf 4}_s,{{\bf 2}}_s)(+1_s,0_s,0_s,0_s)_L]_{5_s 5_s}$  \\
  &$\chi^s_k$ &
 $3\times [({\bf 1}_s,{{\bf 1}}_s)(0_s,+1_s,0_s,-1_s)_L]_{5_s 5_s}$  \\ 
 &${\widetilde \chi}^s_k$ &
 $3\times [({\bf 1}_s,{{\bf 1}}_s)(0_s,0_s,+1_s,+1_s)_L]_{5_s 5_s}$  \\ 
  &$P^s$ &
 $[({\overline {\bf 4}},{\bf 1};{\overline {\bf 4}}_s,{{\bf 1}}_s)(-1,0,0,0;-1_s,0_s,0_s,0_s)_L]_{95_s}$  \\
 &$p^s$ &
 $[({\bf 1},{\bf 1};{\bf 1}_s,{{\bf 1}}_s)(0,-1,0,0;0_s,0_s,-1_s,0_s)_L]_{95_s}$  \\
  &${\widetilde p}^s$ &
 $[({\bf 1},{\bf 1};{\bf 1}_s,{{\bf 1}}_s)(0,0,-1,0;0_s,-1_s,0_s,0_s)_L]_{95_s}$  \\ 
 &$Q^s$ &
 $[({\bf 4},{\bf 1};{\bf 1}_s,{{\bf 2}}_s)(+1,0,0,0;0_s,0_s,0_s,0_s)_L]_{95_s}$  \\  
  &$q^s$ &
 $[({\bf 1},{\bf 1};{\bf 1}_s,{{\bf 1}}_s)(0,+1,0,0;0_s,0_s,0_s,-1_s)_L]_{95_s}$  \\
 &${\widetilde q}^s$ &
 $[({\bf 1},{\bf 1};{\bf 1}_s,{{\bf 1}}_s)(0,0,+1,0;0_s,0_s,0_s,+1_s)_L]_{95_s}$  \\
  &$R^s$ &
 $[({\bf 1},{{\bf 2}};{\bf 4}_s,{\bf 1}_s)(0,0,0,0;+1_s,0_s,0_s,0_s)_L]_{95_s}$  \\
  &$r^s$ &
 $[({\bf 1},{{\bf 1}};{\bf 1}_s,{\bf 1}_s)(0,0,0,-1;0_s,+1_s,0_s,0_s)_L]_{95_s}$  \\ 
 &${\widetilde r}^s$ &
 $[({\bf 1},{{\bf 1}};{\bf 1}_s,{\bf 1}_s)(0,0,0,+1;0_s,0_s,+1_s,0_s)_L]_{95_s}$  \\ 
 &$P^{s s^\prime}$ &
 $[({\overline {\bf 4}}_s,{\bf 1}_s;{\overline {\bf 4}}_{s^\prime},{{\bf 1}}_{s^\prime})(-1_s,0_s,0_s,0_s;-1_{s^\prime},0_{s^\prime},0_{s^\prime},0_{s^\prime})_L]_{5_s 5_{s^\prime}
 }$  \\
 &$p^{s s^\prime}$ &
 $[({\bf 1}_s,{\bf 1}_s;{\bf 1}_{s^\prime},{{\bf
  1}}_{s^\prime})(0_s,-1_s,0_s,0_s;0_{s^\prime},0_{s^\prime},-1_{s^\prime},
 0_{s^\prime})_L]_{5_s 5_{s^\prime}}$  \\
 &${\widetilde p}^{ss^\prime}$ &
 $[({\bf 1}_s,{\bf 1}_s;{\bf 1}_{s^\prime},{{\bf 1}}_{s^\prime})(0_s,0_s,-1_s,0_s;0_{s^\prime},-1_{s^\prime},0_{s^\prime},0_{s^\prime})_L]_{5_s 5_{s^\prime}}$  \\ 
  &$Q^{s s^\prime}$ &
 $[({\bf 4}_s,{\bf 1}_s;{\bf 1}_{s^\prime},{{\bf 2}}_{s^\prime})(+1_s,0_s,0_s,0_s;0_{s^\prime},0_{s^\prime},0_{s^\prime},0_{s^\prime})_L]_{5_s 5_{s^\prime}}$  \\  
   &$q^{s s^\prime}$ &
 $[({\bf 1}_s,{\bf 1}_s;{\bf 1}_{s^\prime},{{\bf 1}}_{s^\prime})(0_s,+1_s,0_s,0_s;0_{s^\prime},0_{s^\prime},0_{s^\prime},-1_{s^\prime})_L]_{5_s 5_{s^\prime}}$  \\
  &${\widetilde q}^{s s^\prime}$ &
 $[({\bf 1}_s,{\bf 1}_s;{\bf 1}_{s^\prime},{{\bf 1}}_{s^\prime})(0_s,0_s,+1_s,0_s;0_{s^\prime},0_{s^\prime},0_{s^\prime},+1_{s^\prime})_L]_{5_s 5_{s^\prime}}$  \\
   &$R^{s s^\prime}$ &
 $[({\bf 1}_s,{{\bf 2}}_s;{\bf 4}_{s^\prime},{\bf 1}_{s^\prime})(0_s,0_s,0_s,0_s;+1_{s^\prime},0_{s^\prime},0_{s^\prime},0_{s^\prime})_L]_{5_s 5_{s^\prime}}$  \\
   & $r^{s s^\prime}$ &
 $[({\bf 1}_s,{{\bf 1}}_s;{\bf 1}_{s^\prime},{\bf 1}_{s^\prime})(0_s,0_s,0_s,-1_s;0_{s^\prime},+1_{s^\prime},0_{s^\prime},0_{s^\prime})_L]_{5_s 5_{s^\prime}}$  \\ 
    & ${\widetilde r}^{s s^\prime}$ &
 $[({\bf 1}_s,{{\bf 1}}_s;{\bf 1}_{s^\prime},{\bf 1}_{s^\prime})(0_s,0_s,0_s,+1_s;0_{s^\prime},0_{s^\prime},+1_{s^\prime},0_{s^\prime})_L]_{5_s 5_{s^\prime}}$  \\\hline
\end{tabular}
\caption{The massless open string spectrum of the ${\cal N}=1$ Type I compactification on $T^6/{\bf Z}_2\otimes{\bf Z}_2\otimes {\bf Z}_3$ with a Wilson line.
The $U(1)$ charges are given in parentheses.}
\label{Z6a} 
\end{table}

\end{document}